\documentclass[twocolumn,final,10pt,a4paper,conference]{IEEEtran}
\usepackage[usenames,dvipsnames]{xcolor}
\usepackage{amsmath,amsthm,graphicx,cite}
\usepackage{epsfig,amsfonts}
\usepackage{graphicx,cite,amssymb,amsmath}
\usepackage{color}
\usepackage{enumitem}
\usepackage{dblfloatfix}
\usepackage[nolist]{acronym}
\usepackage{psfrag}
\usepackage[keeplastbox]{flushend}
\usepackage{mathtools}
\mathtoolsset{showonlyrefs}
\usepackage{placeins}

\newcommand{\library}{\mathcal{F}}
\newcommand{\nfiles}{N}

\newcommand{\capacity}{C}

\newcommand{\Pout}{\ensuremath{\mathsf{P_{out}}}}
\newcommand{\Ps}{\ensuremath{\mathsf{P_{succ}}}}
\newcommand{\Pnot}{\ensuremath{\mathsf{P_{nc}}}}

\newcommand{\Pp}{\mathsf{P}}

\newcommand{\Pk}{p_{\Kreq}}

\newcommand{\Zdiff}{\ensuremath{\mathsf{Z}}}
\newcommand{\zd}{\mathsf{z}}

\newcommand{\Kreq}{\ensuremath{\mathsf{K}}}
\newcommand{\kr}{\mathsf{k}}

\newcommand{\libraryf}[1]{\mathsf{f}_{#1}}

\newcommand{\estef}{\textcolor{black}}

\begin{document}
\begin{acronym}

\acro{MBS}{macro base station}
\acro{MNS}{master node station}
\acro{PMF}{probability mass function}
\acro{SBS}{small base  station}
\acro{BiB}{balls into bins}
\acro{RaP}{random placement}
\acro{rv} {random variable}
\acro{MoP}{most popular placement}
\acro{S}{satellite}
\acro{R}{relay}
\end{acronym}

\begin{acronym}

\acro{MBS}{macro base station}
\acro{MNS}{master node station}
\acro{PMF}{probability mass function}
\acro{SBS}{small base  station}
\acro{BiB}{balls into bins}
\acro{RaP}{random placement}
\acro{rv} {random variable}
\acro{MoP}{most popular placement}
\acro{S}{satellite}
\acro{R}{relay}
\end{acronym}


\title{     Caching  at the Edge: Outage Probability}

\author{
   \IEEEauthorblockN{Estefan\'ia Recayte, Andrea Munari\\
    \IEEEauthorblockA{\IEEEauthorrefmark{1}Institute of Communications and Navigation of DLR (German Aerospace Center),
 \\Wessling, Germany. Email:  \{estefania.recayte, andrea.munari\}@dlr.de}\\
 }
}
\maketitle

\begin{abstract}
Caching at the edge of wireless networks is a key technology to reduce traffic in the backhaul link. However, a concentrated amount of requests during peak-periods may cause the outage of the system, meaning that the network is not able to serve the whole set of demands.  The outage probability is a fundamental metric to take into account during the network design. In this paper,  we derive  the analytical expression of the outage probability as a function of the total amount of users requests, {library size}, requests distribution, cache size and capacity constraints on the backhaul resources. In particular, we focus on a scenario where end-users have no direct connection to the master node which holds the complete library of content that can be requested. A  general formulation of the outage is derived and   studied for two relevant caching schemes, i.e. the random caching scheme and the most popular caching schemes. The exact closed form expressions presented in this paper provide useful insights on how requests, memory and resources can be balanced when the parameters of a cache-enabled network have to   designed.
\end{abstract}

\vspace{1cm}

\section{Introduction}\label{sec:Intro}
The massive increase of multimedia content poses new challenges in wireless networks design. Typical approaches to counteract such enormous capacity demand  consist in increasing spectral resources, i.e. bandwidth, or improving the spatial reuse, i.e. density of transmitters.  However,  in many cases  these techniques may not be applicable due to their inherent costs or complexity. 
At the same time, sparing precious resources represents one of the most important objective for both satellite and terrestrial operators. A promising and   feasible solution which is steadily gaining momentum both in the research and industry community consists in bringing the intended content at the edge of the network by means of caching \cite{magcache}. Indeed, memorizing copies of content close to the users not only alleviates  considerably  the backhaul traffic, but may significantly reduce  latency and power consumption. To achieve this goal, a two-step caching strategy is implemented, pre-fetching the content at the edge (e.g. at small base stations, relays or helpers) during network off-peak periods (\emph{placement phase}) so as to serve the users without consuming backhaul capacity  when the network is congested (\emph{delivery phase}). 

{The effectiveness of caching is driven by a fundamental trade-off concerning the cost-related limits of physical cache.  As a consequence, a proper balance between cache size and resource allocation has to be struck \cite{tradechin}}.  From this standpoint, a meaningful parameter for characterizing the system performance and  which gives an insight of the design layout is given by the outage probability, i.e the probability that a user request cannot be served.  Indeed, once the outage value at which the system should work  is fixed then the memory size can be calculated based on the total  available bandwidth and the total number  of users.  
  
 Based on these considerations, several works have recently investigated outage in caching networks. Interesting results have been obtained in \cite{CaireCachingOutage}, computing the outage probability in device-to-device (D2D) cache enabled-networks where user can download the desired content from a one-hope neighbour. Instead, in  \cite{modelingBastug} the outage probability of a   user  is given in a terrestrial network considering cache-enabled small base stations.  In \cite{HassanCaching,HassanCaching2 } authors derive  a closed form expression of the outage probability for a single user placed at the center of a dense small cell network.    Instead in \cite{hybridZheng}, authors studied  optimization of caching schemes to improve cooperative communications in terms of outage performance gain in a scenario composed by multiple relays and a single user. Considering a multiple amplify-and-forward relay network, the content placement is optimized in \cite{RelayFan} for reducing the outage when relays has unitary cache capacity and  by considering a best relay selection.

This extensive body of research has provided a solid understanding of the potential of caching in serving the request of a specific user, assuming the existence of a connection to both local caches and to nodes keeping copy of all content of interest. On the other hand, scenarios in which multiple users  attempt to retrieve content from the same cache, and cannot rely to a direct backhaul connection have not been tackled yet. 
Such setups are especially relevant in networks (e.g. beyond-5G systems and non-terrestrial networks (NTN)) which foresee a satellite component, employed to deliver content into local caches at ground base-stations. In this case, user terminals are typically not equipped with direct satellite connectivity, and the intermediate tier is responsible to forward content from one end to the other. In this context, only few works have investigated the performance of caching schemes \cite{cachingTef}, and the outage behaviour remains unexplored. Notably, new trade-offs arise, as a proper dimensioning of the satellite link capacity and cache size at ground relays becomes crucial in determining the quality of experience at the end-users.

To bridge this gap, we derive in this paper simple closed-form expressions for the outage probability, considering different statistics for file request distribution, a generic number of users and a  capacity backhaul constraint. In particular, the performance of the network is analysed under two relevant caching schemes, i.e a random caching scheme and the most popular caching scheme. In the former case, we obtain exact expressions, whereas in the latter we overcome the problem complexity deriving a tight approximation of the outage expression which is validated via Monte-Carlo simulations. 
The presented formulations offer interesting insights, which are extensively discussed, and provide a useful design tool. 

The rest of the paper is organized as follows. In Section \ref{sec:sysmodel} the system model is presented while in Section \ref{sec:outageProb} the general expression of the outage is derived. In Section \ref{sec:balls} the outage for each scheme is studied. In Section  \ref{sec:results} the numerical results are given. Finally, Section \ref{sec:Conclusions} addresses the conclusions.

\subsection*{Notation} 
We use sans serif capital letters, e.g. $\mathsf{X}$, for \acp{rv} and their lower case counterparts, e.g.  $\mathsf{x}$, for their realizations. The probability mass function (pmf) of the \ac{rv}  $\mathsf{X}$ is denoted as $\mathsf{P_X}$. Furthermore,  we denote  conditional pmfs as $\Pr\{ \mathsf X = \mathsf x \,|\, \mathsf \mathsf Y = \mathsf y  \} = p_{\mathsf X}(\mathsf x | \mathsf y)$. 

\vspace{.5cm}

\section{System Model}\label{sec:sysmodel}
\begin{figure}[!t]
 \includegraphics[width=0.45\textwidth]{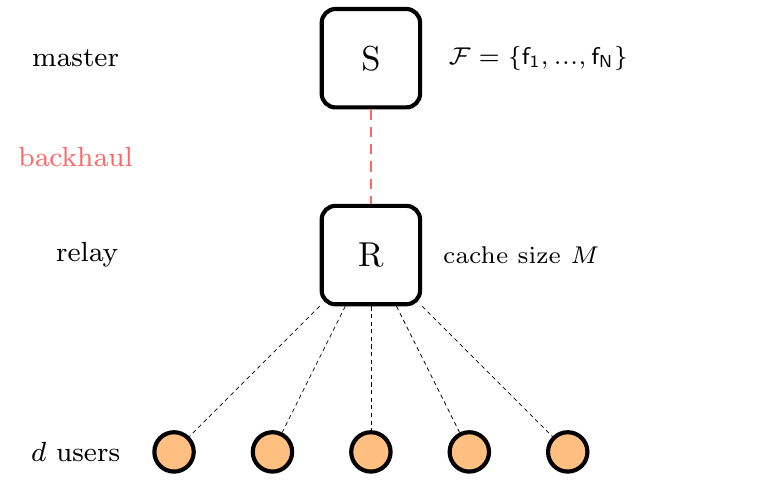}
 \centering \caption{Reference system topology: $d$ users/terminals are connected to the relay R with cache size $M$ files, the relay is connected to the master node S through the backhaul link.   S holds the whole library $\mathcal{F}$ of files. }
 \label{fig:systemMo}
\end{figure}

We consider a two-tier heterogeneous network where end-users are served by a cache-enabled node, which, in turn, is directly connected to a master node.
While this setup applies to different network configurations, we will take as reference throughout our discussion the satellite topology illustrated in Fig.~\ref{fig:systemMo}. Here, a \ac{S} holds a whole library ${\library =\{ \libraryf{1}, \cdots, \libraryf{\nfiles} \}}$ of equal size files. On the ground, a cache-enabled \ac{R} is connected via a backhaul link to \ac{S}, and provides connectivity to users (or terminals) within its cell. Due to memory limitations, only a subset of $M \leq \nfiles$ files can be stored by \ac{R}. Moreover, as typical in current satellite-aided terrestrial networks, we assume that no direct link between users and \ac{S} is available.

In such configuration, let $d$ indicate the number of terminals that concurrently request content from the library, each independently picking a file to download. The requests are processed at \ac{R}, which directly delivers  files present in its cache, and retrieves via the backhaul link content which is not locally available. Aiming to characterise the trade-offs among memory size, backhaul dimensioning and content caching strategies, we assume that enough bandwidth is provided to correctly serve all users-to-relay connections, whereas a limited capacity is available on the relay-to-satellite link. Specifically, we denote the latter quantity by $\capacity$, defined as the maximum number of \emph{different} files that can be retrieved by \ac{R} when attempting to serve users' requests.

Following this notation, the system is said to be in \emph{outage} if the network cannot deliver content to all the $d$ terminals, i.e., if the amount of content that has to be served through the backhaul link exceeds {the capacity constraint} $\capacity$. It is worth noting that the event is driven not only by the available capacity, but also by how the relay caches files based on users demands. To explore this dimension, we consider two well-known and widely employed caching schemes, namely \ac{RaP}  and \ac{MoP}, which are explained  next.

\subsection*{Random placement caching scheme (RaP)}
In the  \ac{RaP} caching scheme, the request distribution is described as follows
\begin{align} \label{puniform}
    p = \frac{1}{N},
\end{align}
{i.e., each file belonging to $\library$ is assumed to be requested with the same probability $p$.}
Accordingly, during the placement phase \ac{R} caches $M$  files from the library uniformly at random, so that the probability for a requested file $\libraryf{i}$ to be present in cache is
\begin{equation} \label{placemop}
\Pr\{\libraryf{i} \text{ is cached in \ac{RaP}}\} = \frac{M}{N} \quad \forall i.
\end{equation}
{The RaP scheme represents a benchmark study case. The analysis of such approach is important, for instance, in scenarios where the actual file requests  distribution is unknown.}

\subsection*{Most popular placement caching scheme (MoP)}
In the \ac{MoP} caching scheme a file $\libraryf{i}$ is requested with probability $p_i$ which follows a Zipf distribution \cite{zipf} with shape parameter $\alpha$ such that
\begin{align}\label{eq:zipf}
p_i = \frac{1}{\beta} \,{i^{-\alpha}} \quad \quad i=1, ..., N
\end{align}
where $\beta := \sum_{j=1}^N j^{-\alpha}$ and the file-index $i$ represents the order based on its popularity.
During the placement phase,    \ac{R} caches the $M$ most probable  files of the library.
The probability that   file $\libraryf{i}$ is cached in the most popular scheme is then

\begin{align}\label{placezipf}
\Pr\{\libraryf{i} \text{ is cached in \ac{MoP}}  \}=
  \begin{cases}
  1 \quad \quad   \quad \quad  i \leq M \\
  0 \quad \quad  M < i \leq N.
  \end{cases}
\end{align}

 \begin{figure*}[t!]
\normalsize
\setcounter{equation}{6}
\begin{align}
\begin{split}
\Pout   =  1- \Bigg[\sum_{\kr=0}^C {d \choose \kr} \, \Pp_\mathsf{nc}^{\kr} \, (1-\Pnot)^{d-\kr}   + \sum_{\kr=C+1}^d {d \choose \kr}\Pp_\mathsf{nc}^{\kr} \, (1-\Pnot)^{d-\kr}  \sum\limits_{\zd=1}^C p_{\Zdiff}(\zd |\kr)  \,   \Bigg] .
\end{split}
 \label{Poutgen}
\end{align}
\hrulefill
\end{figure*}

\begin{figure}[!t]
 \includegraphics[width=0.5\textwidth]{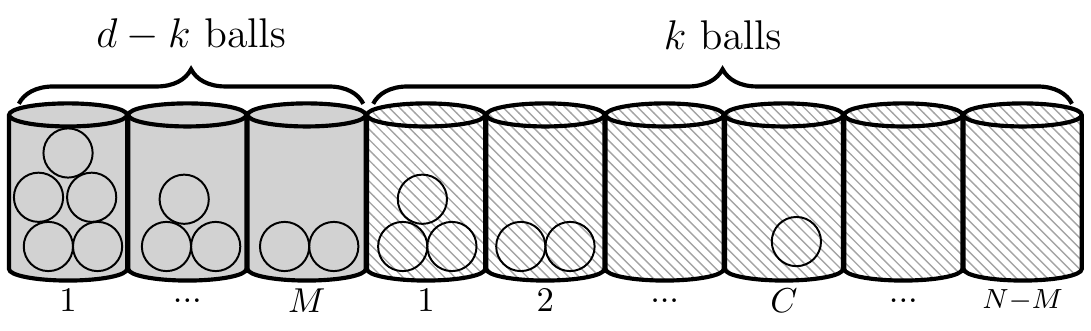}
 \centering \caption{Outage probability represented as  the  balls into bins problem. The $M$ gray colored bins contain   $d-\kr$ requested files which are present in cache (subset $\mathcal{A}$). The $N-M$ dashed bins contain $\kr$ requested files which are not present in cache (subset $\mathcal{B}$). The successful  probability  lies on calculating the probability of having at most $C$ non empty bins in $\mathcal{B}$. }
 \label{fig:ballsinbins}
\end{figure}

\vspace{.5cm}

\section{Outage Probability Formulation} \label{sec:outageProb}

To derive the outage behaviour of the system, we conveniently focus on the complementary quantity \mbox{$\Ps = 1- \Pout$}, capturing  the probability for the network to succeed in serving all users' requests, i.e. that the amount of content to be retrieved via the backhaul does not exceed the capacity constraint $C$.\footnote{We implicitly restrict our attention to the only relevant case $d>C$. For $d\leq C$, in fact, no outage occurs.}

More formally, let us indicate as $\Zdiff$ the rv describing the number of different requested files which are not present in cache. The rv has alphabet $\{0, \cdots, \min(d, N-M)\}$, and allows to readily write
 \begin{equation}\label{eq:Ps}\setcounter{equation}{2}
     \Ps =  \Pr\{\Zdiff  \leq C\}.
 \end{equation}
Let us furthermore introduce the rv $\Kreq$, counting the number of \emph{users} which have picked a file not present in cache. Note that the rv has alphabet $\{0, \dots, d\}$, and that $\Zdiff \leq \Kreq$, since multiple users might ask for the same content. Leaning on this, the expression in \eqref{eq:Ps} can be obtained via the law of total probability as
 \begin{equation}\label{eq:Ps2}
      \Ps = \sum_{\kr=0}^d \Pr\{\Zdiff  \leq C \,| \,\Kreq= \kr\} \Pk(\kr)
 \end{equation}
In turn, the summation in \eqref{eq:Ps2} can be split into two addends. Indeed, whenever \Kreq\ is lower than $C$, all the users can be served with success. Instead, when the number of terminals that request content not in cache is larger than the capacity constraint, the system succeeds only if the amount of  distinct files requested does not exceed $C$ (i.e., if two or more of such users have picked the same file). Applying these remarks, we then have
 \begin{equation}\label{eq:Psucc}
  \Ps = \sum_{\kr=0}^C \Pk(\kr)    + \sum_{\kr=C+1}^d \Pk(\kr) \sum_{\zd =1}^C p_{\Zdiff}(\zd  | \kr).
 \end{equation}

Let us now focus on \Kreq, and denote by $\Pnot$ the probability that a terminal selects a file not present in cache. Recalling that each user independently selects content, and that files are pre-fetched into the relay's cache, the r.v. follows a binomial distribution, i.e. $\Kreq \sim \text{Bin}(d,\Pnot)$, and
\begin{equation}
\Pk(\kr) = {d \choose \kr} \,\Pp_\mathsf{nc}^{\kr} \, (1 {-}\Pnot)^{d-\kr}. \label{pdk}
\end{equation}

Plugging \eqref{pdk} into \eqref{eq:Psucc} finally leads to the general expression for the outage probability reported in \eqref{Poutgen}  at the top of the page. The formulation in \eqref{Poutgen} is handy, as it captures the behaviour of the system under a general caching strategy. In turn, $\Pnot$ and the conditional pmf $p_{\Zdiff}(\zd  | \kr)$ are specific to the implemented content storage policy, and will be derived in details in the next section for both the MoP and RaP approaches.


\vspace{.5cm}

\section{Balls into bins problem applied to caching} \label{sec:balls}
In order to instantiate the calculation of the outage probability for the considered caching strategies, it is convenient to map our setting onto a \acl{BiB} (BiB) setup. The general \ac{BiB} problem, see e.g. \cite{Kotz}, consists in independently throwing $d$ balls into $N$ bins. As illustrated in  Fig.~\ref{fig:ballsinbins}, this can be cast to our case by having each bin associated to a file of the library, and by having balls represent user requests. Following this parallel, the possibility for more balls to land into the same bin corresponds to having multiple users asking for a common library element.

Without loss of generality, we split the bins into two subsets, labelled $\mathcal A$ and $\mathcal B$. The first has cardinality $M$, and indicates the files that are cached at \ac{R}, while the second is composed of the $N-M$ bins that denote files only available via the backhaul link. Recalling the notation of Sec.~\ref{sec:outageProb}, a ball will then land into a bin of the two classes with probability $1-\Pnot$ and $\Pnot$, respectively, and the rv \Kreq\ counts the number of balls thrown onto bins in the second category. Furthermore, the pmf $p_{\Zdiff}(\zd |\kr)$ can conveniently be seen as describing the number non-empty bins in $\mathcal B$ after $d$ throws have been performed, conditioned on having $\Kreq$ balls land into bins belonging to $\mathcal B$.
Notably, as will be discussed in the following, exact closed-forms for such distribution can be derived when bins are picked uniformly, whereas tight approximations can be obtained when balls have different landing probabilities.





\begin{figure*}[t!]
\normalsize
\setcounter{equation}{9}
\begin{align}
\begin{split}
\Pp^{\mathsf{(RaP)}}_{\mathsf{out}} =  1- \Bigg[\sum_{\kr=0}^C {d \choose \kr} \,  \frac{ M^{d-\kr}}{N^{d}}     (M-N)^\kr + \sum_{\kr=C+1}^d  {d \choose \kr}  \sum\limits_{\zd=1}^C \frac{M^{d-\kr}}{N^d} {N-M \choose \zd} \,\mathit{S}(\kr,\zd)\, \zd!\Bigg].
\end{split}
 \label{Poutrap}
 \end{align}
\hrulefill
\end{figure*}

\subsection{Random placement caching scheme}

Following the \ac{BiB} parallel, the \ac{RaP} scheme  corresponds  to assuming that each ball is thrown uniformly at random over the available bins. The probability that exactly $\Zdiff = \zd$ bins out of the $N-M$ in class $\mathcal B$ are non empty given that $\Kreq = \kr$ launches have landed there can then be written as
 \begin{align}\setcounter{equation}{7}
 p_{\Zdiff}^{\mathsf{(RaP)}}(\zd|\kr)  = \frac{{N-M \choose \zd} \,\mathit{S}(\kr,\zd)\, \zd!}{(N-M)^\kr}\label{eq:pdelta}
 \end{align}
where \[\mathit{S}(\kr,\zd) = \frac{1}{\zd !} \sum_{j=0}^\zd (-1)^j \binom{\zd}{j}  (\zd-j)^\kr.\] 
The result follows by counting  the favorable cases over all the possibles outcomes. In particular, we observe that there are ${N-M \choose \zd}$ ways of choosing $\zd$ files from $N-M$. For each such case, $\mathit{S}(\kr,\zd)$, denoting the Stirling number of the second kind \cite{stirling}, counts  all the possible of ways in which $\kr$ users can request for $\zd$  different files. Finally, $\zd!$  accounts for all the possible permutations of $\mathit{S}(\kr,\zd)$.

In order to compute the distribution of $\Kreq$, on the other hand, we observe that with the RaP policy a user requests a content that was not cached with probability
\begin{equation}\label{Pncrap}
\Pp_{\mathsf{nc}}^{\mathsf{(RaP)}} = 1- \frac{M}{N}
\end{equation}
so that, from \eqref{pdk}, 
\begin{equation}\label{eq:pkrap}
\Pk^{(\mathsf{RaP})}(\kr) = {d \choose \kr} \,\Big(1-\frac{M}{N}\Big)^{\kr} \, \Big(\frac{M}{N}\Big)^{d-\kr}.  
\end{equation} 

Leaning on these results, the outage probability $\Pp_{\mathsf{out}}^{\mathsf{(RAP)}}$ can be derived by plugging \eqref{eq:pdelta} and \eqref{Pncrap} into \eqref{Poutgen}, obtaining after some simple  
manipulations the final expression given in \eqref{Poutrap} at the top of the page.

\subsection{Most popular placement caching scheme}

In the MoP setup, files, i.e. bins, are chosen with different probability. To approach the problem, let us again condition our observation on having $\Kreq = \kr$ terminals selecting contents that are not present in cache. Under this assumption, we can focus on a simpler \ac{BiB} problem, where $\kr$ throws are performed, and each ball can fall solely onto one of the $N-M$ bins in $\mathcal B$. Specifically, recalling the Zipf distribution reported in \eqref{eq:zipf}, the $t-$th bin in this problem is chosen with probability
\begin{align}
    q_t & =   \frac{(M+t)^{-\alpha}}{\sum_{i=M+1}^N i^{-\alpha}} \quad     t = 1, ..., N-M\label{eq:qt}
\end{align}
where $\sum\nolimits_{t=1}^{N-M} q_t = 1$.

In this setup, the derivation of the probability to have \mbox{$\Zdiff = \zd$} not empty bins, i.e., the sought $p_{\Zdiff}(\zd|\kr)$, is known as the \emph{occupancy problem}, for which, despite the simple conceptual formulation, a close-form solution is still elusive. To capture the performance of our system we thus recur to the approximation proposed in \cite{Kotz}, and write the pmf of the number of \estef{non} empty bins as
\begin{align}\label{plambda}\setcounter{equation}{10}
p_{\Zdiff}(\zd|\kr)\approx   \frac{1}{\sqrt{2 \pi \sigma_{\kr }^2}}\exp\Big\{-\frac{(\zd-\mu_{\kr })^2 }{2 \sigma_{\kr}^2}\Big\}
\end{align}
where
\begin{align}
\begin{split}
\mu_{\kr } & := (N-M) -\sum_{t = 1}^{N-M} e^{-\kr q_t} \\
 \sigma_{\kr}^2& := \sum_{t=1}^{N-M} e^{-\kr q_t} \big(1 - e^{-\kr q_t}\big) -\frac{1}{\kr}\Big(\sum_{t=1}^{N-M} \kr e^{-\kr q_t}q_t\Big)^2.
\end{split}
\label{musigma}
\end{align}
\estef{Referring to our caching problem, the equation given in \eqref{plambda} indicates the probability that $\mathsf{k}$  terminals request  for $\zd$ different files which are not cached. }

 \begin{figure}[t]
 \includegraphics[width=0.5\textwidth]{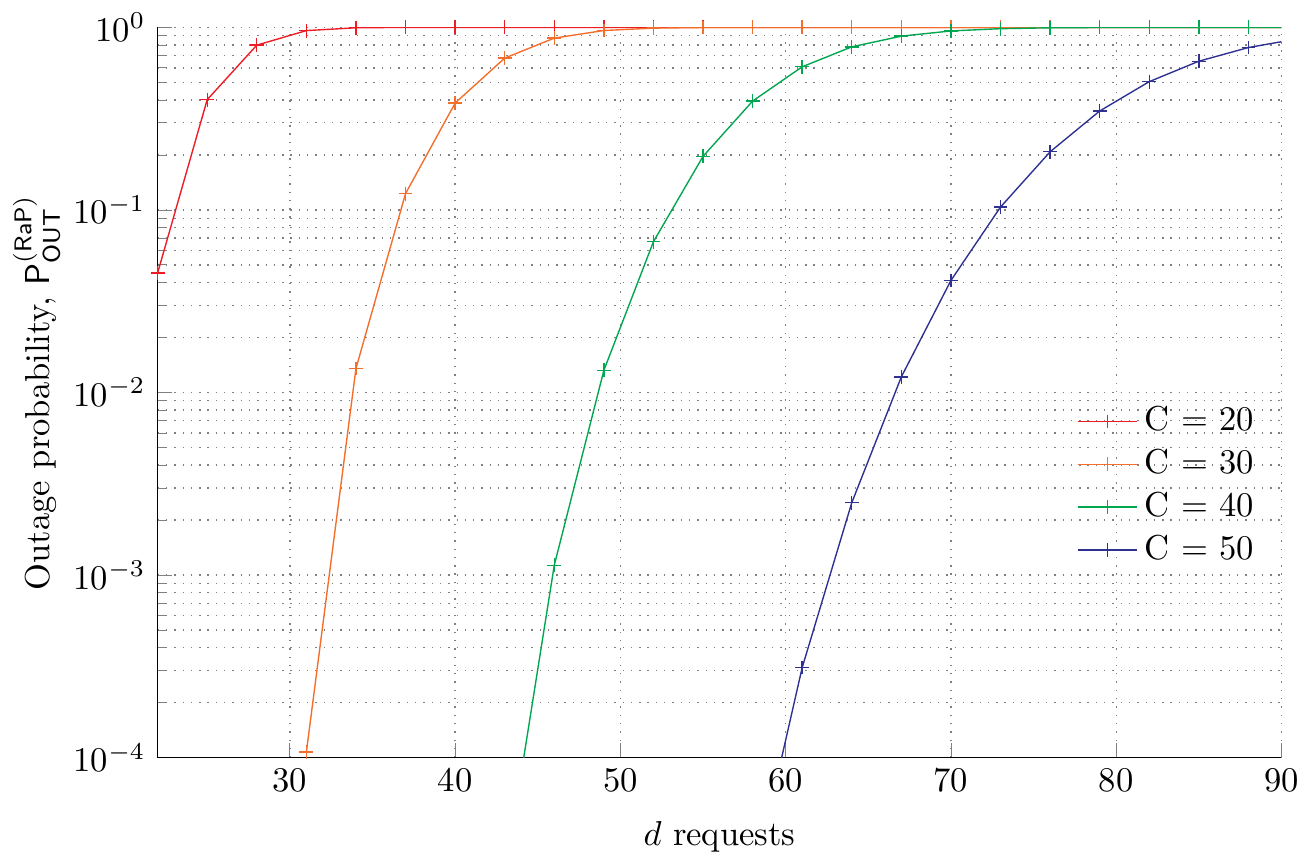}
 \centering \caption{Outage probability in the random placement scheme in function of the number of request $d$  for different capacity levels   ${C=20,30,40, 50}$ when  memory size $M = 10$,  and library size  $N=100$.    } \vspace{-.6cm}
 \label{fig:p_unif}
\end{figure}

Finally, we derive also for the MoP case the probability to request a non-cached file, i.e.  $\Pp^{\mathsf{(MoP)}}_{\mathsf{nc}}$. Leaning on \eqref{eq:zipf} and recalling the caching policy in \eqref{placezipf}, we obtain
\begin{align}\label{pncmop}
\Pp^{\mathsf{(MoP)}}_{\mathsf{nc}} & =  1 -\sum_{i=1}^M p_i \\
& = 1- \frac{1}{\beta}\sum_{i=1}^M\frac{1}{i^{\alpha}} = \frac{1}{\beta}\sum_{i=M+1}^N\frac{1}{i^{\alpha}}.
\end{align}
From \eqref{pncmop}, the binomial pmf of $\mathsf{K}$ follows then as
\begin{align}\label{eq:pkmop}
  \Pk^{(\mathsf{MoP})}(\kr)  
  = & \frac{1}{\beta^d}  {d \choose \kr}  \left(\sum_{i=M+1}^N\frac{1}{i^{\alpha}}\right)^{\kr} \, \Big(\sum_{i=1}^M\frac{1}{i^{\alpha}}\Big)^{d-\kr}\end{align}
A good-approximated expression of outage probability $\Pp^{\mathsf{(MoP)}}_{\mathsf{out}}$ in the most popular placement caching scheme  is obtained by inserting \eqref{plambda} and \eqref{pncmop} into \eqref{Poutgen}. After simple manipulations we eventually obtain \eqref{Poutmop}, reported at the top of next page.

\begin{figure*}[t!]
\normalsize
\setcounter{equation}{12}
\begin{align}
\begin{split}
\Pp^{\mathsf{(MoP)}}_{\mathsf{out}}  \approx  1- \frac{1}{\beta^d}\Bigg[\sum_{\kr=0}^C {d \choose \kr}  \Big(\sum_{i=M+1}^N\frac{1}{i^{\alpha}} \Big)^{\kr} \Big(\sum_{i=1}^M\frac{1}{i^{\alpha}}\Big)^{d-\kr}   + \sum_{\kr=C+1}^d {d \choose \kr} \sum\limits_{\zd=1}^C \frac{1}{\sqrt{2 \pi \sigma_{\kr }^2}} e^{-\frac{1}{2  \sigma_{\kr}^2}(\zd-\mu_{\kr })^2 } \Big(\sum_{i=M+1}^N\frac{1}{i^{\alpha}}\Big)^{\kr}  \Big(\sum_{i=1}^M\frac{1}{i^{\alpha}}\Big)^{d-\kr}   \Bigg] .
\end{split}
 \label{Poutmop}
 \end{align}
\hrulefill
\end{figure*}

 \section{Results}\label{sec:results}

 In our first scenario, we assume a random placement in the cache. Users are connected to a relay with cache size ${M = 10}$ files, while the library cardinality is $N=100$.   In Fig.~\ref{fig:p_unif} the outage probability of RaP as a function of the number of requests for different values of backhaul capacity $C$ is plotted.  As expected,  given $d$ requests the outage probability decreases by increasing the backhaul capacity, since a larger number of requests can be served. However,  this caching scheme requires high  backhaul capacity for operating a relatively low levels of outage.  For instance, the network demands a capacity  $C=40$ for ensure that  simultaneously $d=50$  requests are served with $\Pp_{\mathsf{out}}^{\mathsf{(RaP)}} =0.025$. The plot shows us that, in the case of equiprobable files,   the benefit  obtained from a cached network is minimal. As a matter of fact, this cache architecture does not significantly alleviate the traffic in the backhaul and, to operate with relatively low outage,  the network needs to allocate backhaul resources in the order of the number of users that are active in the system.

  \begin{figure}[t]
 \includegraphics[width=0.5\textwidth]{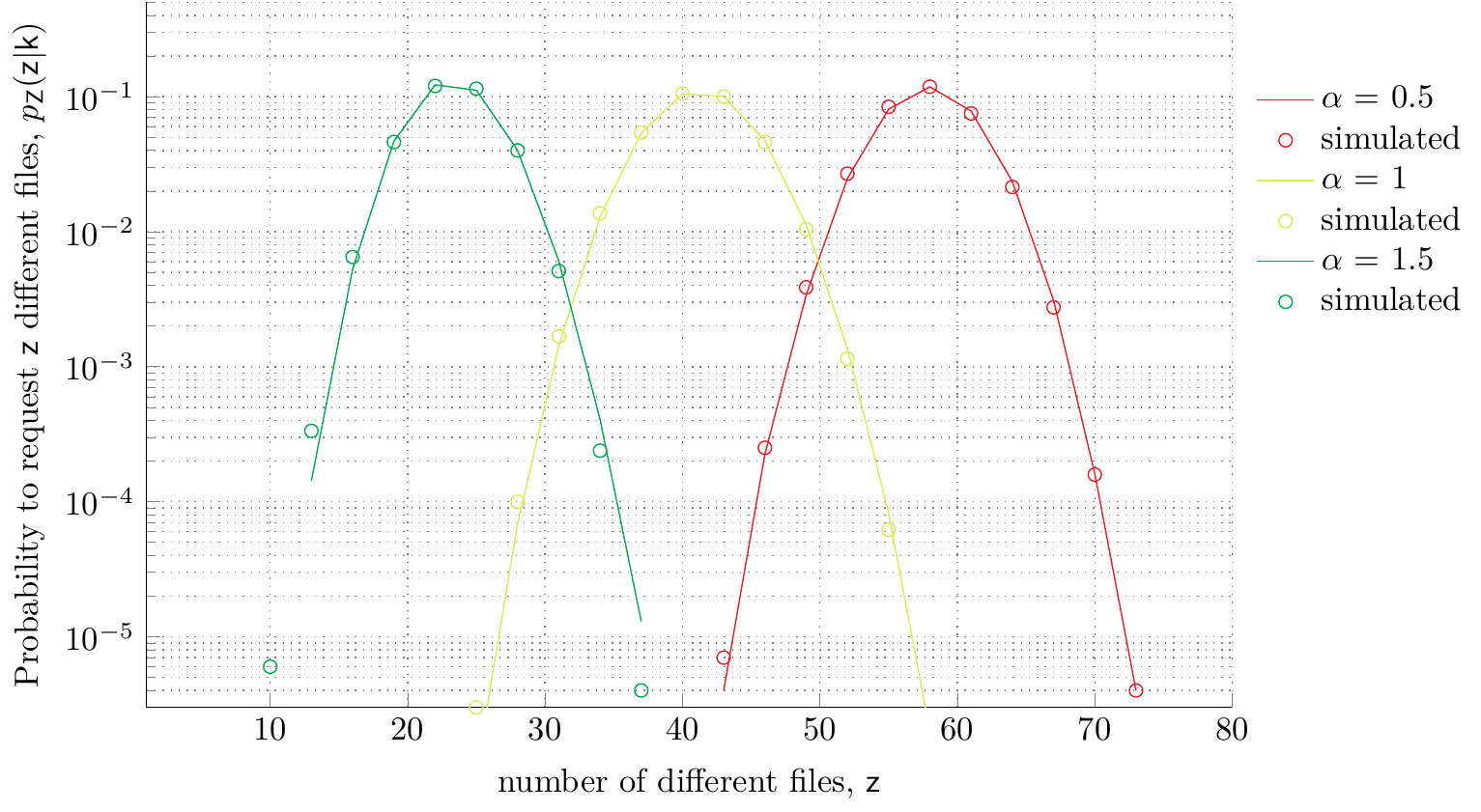}
     \centering \caption{Probability to request $\mathsf{z}$ different files given the demand of  $\mathsf{k=100}$ users when $N = 100$ and files follows the Zipf distribution with parameter $\alpha = 0.5, 1$ and $1.5$. The solid curves represent our analytical approximation in \eqref{plambda} while circles are obtained via Monte-Carlo simulations.}
 \label{fig:pzz}
\end{figure}

 \begin{figure}[t]
 \includegraphics[width=0.5\textwidth]{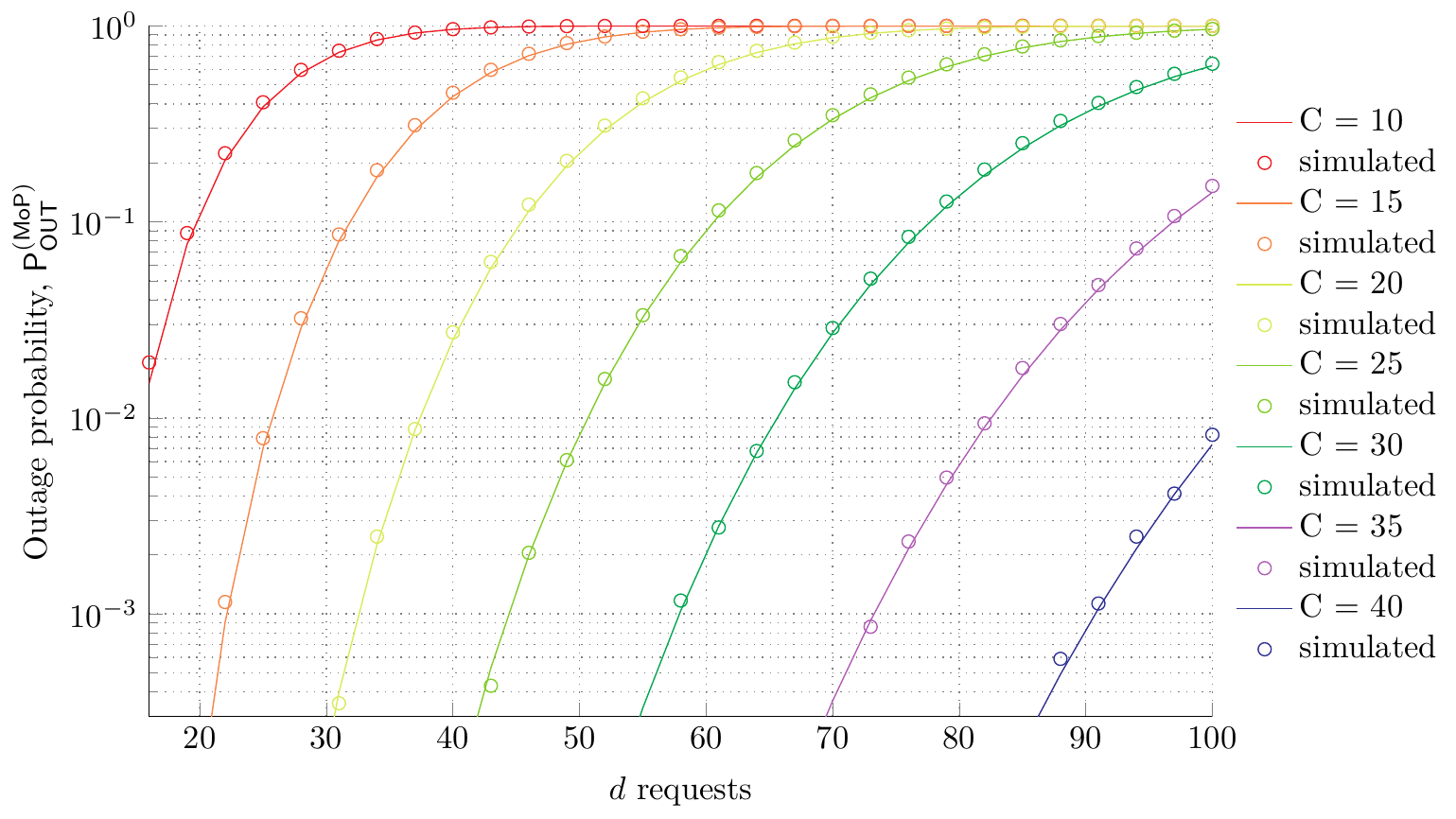}
 \centering \caption{Outage probability in the most popular caching scheme in function of the number of request $d$  for different capacity levels   ${C=10,15,20,25,30,35,40}$ when   $M = 10$, $\alpha=1$ and $N=100$. Given a constraint capacity $C$, the solid lines indicate  the results obtained with our analytical approximation while dot  markers indicate the corresponding result obtained by Monte-Carlo.}
 \label{fig:p_zipf}
\end{figure}

\estef{Let us assume an special case of  the MoP  where $M=0$, i.e. non of the files are cached and users request for content according to the Zipf distribution given in \eqref{eq:zipf}. Under this assumption, $p_{\Zdiff}(\zd|\kr)$ indicates the probability that $k$ users demand for $\mathsf{z}$ different files. In Fig.~\ref{fig:pzz} the probability that $\mathsf{k}=100$ users request  for $\mathsf{z}$ different files is plot for three different Zipf parameters, i.e. $\alpha$, when the library size is  $N =100$. The solid curves represent  the analytical approximation obtained in \eqref{plambda} while circles indicates the  results obtained via Monte-Carlo. The plot shows the tightness and validity of the approximation for different values of $\alpha$. When the skewness of the distribution is higher, i.e. $\alpha = 1.5$, user requests are concentrated in few files as shown by the green curve. In fact, for the considered Zipf parameter in mean   $\mu_{100}= 23.36$ different files are requested. Instead, if we consider a lower Zipf parameter, for instance $\alpha= 0.5$, we observe that requests are spread over a larger number of files and in  mean   $\mu_{100}= 57.79$ different files $z$ are requested.  }

 In Fig.~\ref{fig:p_zipf} the probability of outage as a function of the number of requests is plotted for the most popular placement caching scheme for different values of backhaul capacity. In the figure, solid lines report the analytical approximation, while markers the outcome of Montecarlo simulations. The reported trends were obtained assuming a shape parameter of the Zipf distribution $\alpha =1$, ($\alpha$ typically assumes values in $[0.5,1.5]$, see e.g.  \cite{zipf}). Moreover, the library size is $N=100$ while the memory size $M=10$, allowing a direct comparison with the RaP performance discussed earlier. As a first remark, we observe that the analytical results offer a very tight match to the simulations, prompting how the derived equations provide a simple yet effective tool for a preliminary system design. Furthermore, the plot shows the efficiency of the caching scheme due to the fact that more requests are concentrated in a small number of files. By increasing the backhaul capacity, significant gains in terms of number of requests served for a fixed outage probability is observed. For instance, a network operating  at  $\Pp_{\mathsf{out}}^{\mathsf{(MoP)}} = 0.02$
 can serve simultaneously $d=27$ with only requiring a capacity $C=15$. As soon as we double the backhaul capacity, i.e. $C=30$, then $d=68$ users can be served by ensuring the same outage probability. Unlike the \ac{RaP} approach, the \ac{MoP}  caching scheme  provides a huge gain of resources given that most of the requests are served by cached content.

\begin{figure}[t]
 \includegraphics[width=0.5\textwidth]{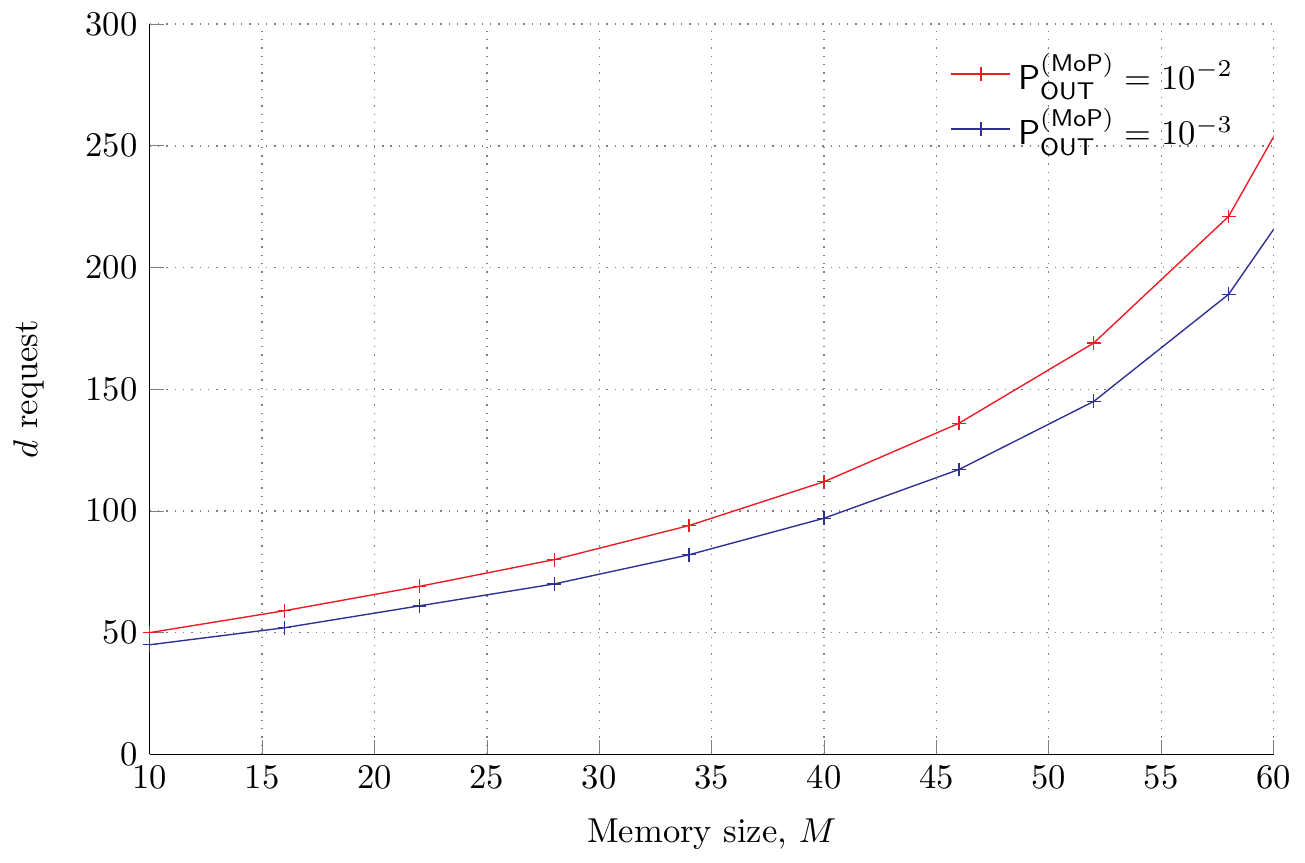}
 \centering \caption{Number of request that the system can support in the  most popular placement scheme given in MoP ${\Pp_{\mathsf{out}}  = 10^{-2}}$ and ${\Pp_{\mathsf{out}}  = 10^{-3}}$ in function of the memory size when ${C = 30}, {\alpha =0.8},{ N  =100.}$}
 \label{fig:p_md}
\end{figure}

The results obtained so far can be applied to design a cache network under fixed requirements. A relevant example could be when a operator has to decide the cache dimension given a constraint on the backhaul capacity while  warranting a certain outage probability $\Pout$. Based on \eqref{Poutmop},   the number of users successfully served by the network can be derived as a function of the memory size given a  $\Pp_{\mathsf{out}}$ of MoP. Thus, in Fig.~\ref{fig:p_md} we show, for $\Pp_{\mathsf{out}}^{\mathsf{{(MoP)}}}   =10^{-2}$ and for $\Pp_{\mathsf{out}}^{\mathsf{{(MoP)}}}  =10^{-3}$, the number of users that can be served $d$ as a function of the memory size $M$ when the
capacity $C=30$, the library size $N=100$ and the Zipf parameter $\alpha  =0.8$. The maximum number of users that can be simultaneously served can be determined  from the plot by  choosing a $\Pp_{\mathsf{out}}^\mathsf{(MoP)}$   and fixing the memory size $M$.

Furthermore, we want to highlight that the results obtained for the MoP are also valid in a scenario where multiple relays are consider as long as relays have the same memory size.
In fact it is easy to check that derivation does not change.

\section{Conclusions}\label{sec:Conclusions}
In this work we consider an heterogeneous network with cache capability and  
we derive the outage probability when multiple users demand  for content. In particular two caching schemes were considered. We derived a closed-form expression of the outage probability when a random caching scheme is on place. A well-approximated expression was obtained and then verified via Monte-Carlo for the most popular caching scheme. {The outage probability was derived as a function of the number of total requests $d$, cache size $M$, total number of files $N$, requests distribution $p$ (in case of RaP) or $p_i$ (in case of MoP) and capacity constraint $C$.}
The results provide useful hints at the time of design a cached network. For example, one can have a quick and easy understand in trade-off between backhaul resources and  maximum number of users that can be  served.

 \vspace{.5cm}

\newpage
 \flushend
\bibliographystyle{IEEEtran}
\bibliography{IEEEabrv,references}

\end{document}